\documentclass[11pt]{amsart}

\theoremstyle{definition}

\theoremstyle{remark}

\begin{document}

\title[]{Alternative space-time for the point mass}
\author{L. S. Abrams}
\address{Lockheed-California Company, Burbank, California 91520}

\thanks{Published in Phys. Rev. D {\bf 20}, 2474 (1979).}
\thanks{(Received 11 March 1977; revised manuscript received 30 July 1979)}

\begin{abstract}
Schwarzschild's actual exterior solution $(g_{\rm S})$ is resurrected,
and together with the manifold $M_0=R^4-\{r=0\}$ is shown to
constitute a space-time possessing all the properties historically
thought to be required of a point mass. On the other hand, the
metric $(g_{\rm DW})$ that today is ascribed to Schwarzschild, but
which was in fact first obtained by Droste and Weyl, is shown to
give rise to a space-time that is neither equivalent to
Schwarzschild's nor derivable from the ``historical'' properties
of a point mass. Consequently, the point-mass interpretation of
the Kruskal-Fronsdal space-time $(M_W,g_{\rm KF})$ can no longer be
justified on the basis that it is an extension of Droste and
Weyl's space-time. If such an interpretation is to be maintained,
it can only be done by showing that the properties of $(M_W,g_{\rm KF})$
are more in accord with what a point-mass space-time should
possess than those of $(M_0,g_{\rm S})$. To do this, one must first
explain away three seeming incongruities associated with
$(M_W,g_{\rm KF})$: its global nonstationarity, the two-dimensional
nature of its singularity, and the fact that for a finite interval
of time it has no singularity at all. Finally, some of the
consequences of choosing $(M_0,g_{\rm S})$ as a model of a point mass
are discussed.
\end{abstract}
\maketitle
\section{Notation and introduction}

Let $K$ denote the analytic manifold consisting of $R^4$ together
with the single-chart atlas $\{R^4,Id\}$, and let the thereby-defined
coordinates be denoted by $(t,x,y,z)$. Let $L$ denote the line
$x=y=z=0$, and $M_0$ the submanifold $K-L$. Since it is essential
to what follows, note that the dimensionality of any point set on
a manifold is determined by the set's description in terms of the
admissible coordinates, and thus has nothing to do with the
subsequent choice of a metric. Thus, e.g., the assertions that $L$
is a line, and that its intersection with $t=$ const is a point,
are valid no matter what metric may be assigned to $M_0$.\par
The properties that the metric $(g)$ of a single, nonrotating,
nonradiating, uncharged point mass\footnote{Henceforth, this
description will be shortened to point mass.} should possess were
first stated by Einstein \cite{Einstein15} in 1915, and
subsequently employed by Schwarzschild \cite{Schw16a}, Droste
\cite{Droste17}, Weyl \cite{Weyl17}, Hilbert
\cite{Hilbert24}, etc., to derive the exact form of $g$. An
explicit formulation of all these properties, including those that
were only tacit (e.g., Lorentz-signature, global time coordinate),
together with those of the associated manifold $(M)$ was given by
Finkelstein \cite{Finkelstein58} in 1958. Specifically, $M$
should be $M_0$, while $g$ should be analytic, Lorentz
signature, and
\begin{list}{}{}
\item (a) a solution on $M$ of Einstein's free-space field
equations;
\item (b) invariant under time translations;
\item (c) invariant under spatial rotations;
\item (d) (spatially) asymptotically flat;
\item (e) inextendable to $L$;
\item (f) invariant under spatial reflections;
\item (g) invariant under time reflection;
\item (h) have a global time coordinate.
\end{list}
Contrary to popular opinion, an analytic, Lorentz signature metric
$(g_{\rm S})$ possessing properties (a)-(h) on $M_0$ does exist. Indeed,
it has been available ever since 1916, when Schwarzschild \cite{Schw16a}
derived it. Introducing quasipolar coordinates via
\begin{equation}\label{1}
x=r\sin\theta\cos\phi,~y=r\sin\theta\sin\phi,~z=r\cos\theta
\end{equation}
with
\begin{equation}\label{2}
0<r,~0<\theta<\pi,~0\leq\phi<2\pi,
\end{equation}
the components of $g_{\rm S}$ in these coordinates\footnote{Since
quasipolar coordinates are inadmissible on the surface $x=y=0$, it
is to be understood here and afterwards that whenever a line
element is displayed in terms of such coordinates, this is done
solely for ease of recognition, and that the metric is actually
defined by its components in terms of $(t,x,y,z)$. In the case of
$g_{\rm S}$ these quasi-Cartesian components are analytic everywhere on
$M_0$.} are given by
\begin{equation}\label{3}
ds^2=(1-\alpha/f)dt^2-(1-\alpha/f)^{-1}{f'}^2dr^2-f^2d\Omega^2,
\end{equation}
where $\alpha$ is a positive constant,
\begin{equation}\label{4}
d\Omega^2\equiv d\theta^2+d\phi^2\sin^2\theta,
\end{equation}
\begin{equation}\label{5}
f\equiv(r^3+\alpha^3)^{1/3},
\end{equation}
and the prime denotes differentiation with respect to $r$.\par
That the metric possesses properties (b), (c), (d), (f), (g), and
(h) is evident by inspection. That it possesses property (a) can
be verified by direct substitution into the field equations, or by
examining Schwarzschild's derivation. Lastly, that it has property
(e) follows from the violation\footnote{Let $C_a$ denote the
circle of ``radius'' $a$ defined by $\{(t,r,\theta,\phi)|t=\text{const},
r=a, \theta=\pi/2\}$. The proper circumference of $C_a$ is seen
from (\ref{3}) to be $2\pi f(a)$. As $a\rightarrow 0^+$, this tends
to $2\pi\alpha$, rather than to zero as local flatness requires.}
of local flatness at every point of $L$. Clearly, $(M_0,g_{\rm S})$ does
not contain a black hole.
\section{The Droste-Weyl metric}
In 1917, Droste \cite{Droste17} and, independently,
Weyl, \cite{Weyl17} derived a metric $(g_{\rm DW})$ for the point
mass by techniques similar to that used by Schwarzschild. Letting
$\overline{M}_0$ denote a copy of $M_0$ with the same admissible
coordinates $(t,x,y,z)$ as before, and letting $(t,\bar{r},
\theta,\phi)$ denote the same quasipolar coordinates defined by
(\ref{1}) and (\ref{2}), the components of $g_{\rm DW}$ are defined on
that portion $(\overline{M}_{\alpha})$ of $\overline{M}_0$ for which
$\bar{r}>\alpha$ by
\begin{equation}\label{6}
ds^2=(1-\alpha/\bar{r})dt^2
-(1-\alpha/\bar{r})^{-1}d\bar{r}^2-\bar{r}^2d\Omega^2,
\end{equation}
and on the remainder of ${M}_0$ by analytic continuation. Note
that the $\bar{r}$ in (\ref{6}), like the $r$ in (\ref{3}), is the
ordinary quasipolar coordinate $(x^2+y^2+z^2)^{1/2}$, and thus its
${M}_0$-spanning values satisfy $\bar{r}>0$.\par
Believing that $g_{\rm DW}$ was ``equivalent'' to $g_{\rm S}$, both Droste and
Weyl credited their results to Schwarzschild. A few years later,
Hilbert \cite{Hilbert24} opined that the form of $g_{\rm DW}$ was
preferable to that of $g_{\rm S}$, and ever since then the phrase
``Schwarzschild solution'' has been taken to mean $g_{\rm DW}$ rather
than $g_{\rm S}$.
\section{Inequivalence of the Schwarzschild and Droste-Weyl
space-times}
Let us relabel the $r$ coordinate of points of $M_0$ by means of
the coordinate transformation
\begin{equation}\label{7}
T:~\bar{r}=f(r)=(r^3+\alpha^3)^{1/3},~r>0
\end{equation}
(\footnote{Here too, as in footnote 2, this equation is written in
polar coordinates for simplicity. The actual transformation is
defined by the corresponding relations in terms of the
quasi-Cartesian coordinates $(t,x,y,z)$ and $(t,\bar x,\bar y,\bar
z)$, obtainable from (\ref{7}) by cubing both sides and then
using (\ref{1}) to eliminate the polar coordinates, e.g.,
$\bar{x}^3=x^3+{\alpha}^3x^3/(x^2+y^2+z^2)^{3/2}$, etc. These
relations are defined everywhere on $M_0$.}) so that the
$M_0$-spanning values of $\bar r$ satisfy $\bar r>\alpha$.\par
Under $T$, $g_{\rm S}$ is carried into $\bar g_{\rm S}$:
\begin{equation}\label{8}
ds^2=(1-\alpha/\bar{r})dt^2
-(1-\alpha/\bar{r})^{-1}d\bar{r}^2-\bar{r}^2d\Omega^2.
\end{equation}

In the early years of relativity, when ``physical equivalence'' was
thought to be a relationship between metrics, the formal identity
of (\ref{6}) and (\ref{8}) was taken to mean that $\bar g_{\rm S}$ (and
{\it a fortiori} $g_{\rm S}$) and $g_{\rm DW}$ describe the same physical
phenomenon (which is the reason that both Droste and Weyl
attributed their results to Schwarzschild). Today, however, it is
recognized that physical equivalence is a relationship between
space-times \cite{HE74a}, and that for two space-times
$(M,g_1)$ and $(M,g_2)$ defined on the $same$ manifold to be
equivalent, it is not only necessary that a coordinate
transformation carry $g_1$ into $g_2$, but also that the
transformation be a diffeomorphism \cite{HE74b}. In the
present case this latter requirement is not met, since $T$ carries
$M_0$ {\it in}to $\overline{M}_0$ (viz., onto $\overline{M}_{\alpha}$), rather
than {\it on}to it, and is thus not a diffeomorphism. Consequently
$(M_0,g_{\rm S})$ and $(\overline{M}_0,g_{\rm DW})$ are inequivalent space-times.
This invalidates one of the two historical bases for regarding
$g_{\rm DW}$ as the metric of a point mass.

\section{Derivation of the Droste-Weyl space-time}
The other historical basis for the point-mass interpretation of
$g_{\rm DW}$ is that Droste \cite{Droste17}, Weyl
\cite{Weyl17}, Hilbert \cite{Hilbert24}, etc., believed
that they had derived it by explicitly tailoring the general
metric to have properties (a)-(h). As will now be shown, their
derivations produce not $g_{\rm DW}$, but rather $\bar g_{\rm S}$, the
coordinate-transformed version of $g_{\rm S}$, in which $\bar r>\alpha$
spans $M_0$.\par
Consider a typical derivation\footnote{Although what follows is
really a generic version of the classical derivation, it is
closely approximated by that in Ref. \cite{Weinberg72}.}. Although the
manifold is not specified, there is no basis for supposing that
anything other than $M_0$ was contemplated, so that this will be
assumed in what follows.\par
To begin with, properties (b), (c), (f), (g), and (h) are imposed
on the metric, which as is well known
\cite{Schw16a,Droste17,Weyl17,Hilbert24} restricts
the line element to the form
\begin{equation}\label{9}
ds^2=A(r)dt^2-B(r)dr^2-C(r)d\Omega^2
\end{equation}
in terms of the quasipolar coordinates defined by (\ref{1})
and (\ref{2}). Note that property (h) requires that $A$ be
positive on $M_0$, and that this together with the
Lorentz-signature requirement compels $B$ and $C$ to be positive
as well.\par
Next, the unknown $C$ is eliminated by making the coordinate
transformation
\begin{equation}\label{10}
\bar r=[C(r)]^{1/2},
\end{equation}
thereby transforming (\ref{9}) into
\begin{equation}\label{11}
ds^2=\overline{A}(\bar r)dt^2-\overline{B}(\bar r)d{\bar r}^2
-{\bar r}^2d\Omega^2.
\end{equation}
Property (a) is now imposed, which leads to two ordinary
differential equations for the unknown ${A}$ and ${B}$. The
solution of these equations determines both unknowns, apart from
two constants of integration. Property (d) is used to eliminate
one of them, whereupon the metric, hereinafter referred to as
$g'_{\rm S}$, takes the {\it form} of (\ref{6}). Finally, a comparison
is made with Newtonian physics to evaluate $\alpha$ - for our
purposes, this step can be replaced by imposing property (e),
which requires that $\alpha$ be nonzero.\par As is evident from
this brief sketch, in order for $g'_{\rm S}$ to be $g_{\rm DW}$ it is
necessary that the $M_0$-spanning values of $\bar r$ in (\ref{11})
be the same as those of the $\bar r$ in (\ref{6}), viz., $\bar r>0$
\cite{Brans65}. But as can be seen from (\ref{10}), the
$M_0$-spanning values of $\bar r$ in (\ref{11}) depend on the
behavior of $C(r)$ ar $r$ varies over $(0,\infty)$. Thus the only
way to determine the values in question is to substitute (\ref{9})
directly into the empty-space field equations and solve for the
most general $C(r)$ compatible with properties (d) and (e). This
is done in the Appendix, and the result is
\begin{equation}\label{12}
[C(r)]^{1/2}=\alpha/[1-A(r)],
\end{equation}
where $\alpha$ is a positive constant and $A$ is any analytic,
strictly monotonic increasing function of $r$ which tends to zero
as $r\rightarrow 0$ and behaves like $1-\alpha/r$ as $r\rightarrow
\infty$. Together with (\ref{10}), the just-mentioned properties of $A$
show that the $\bar r$ in (\ref{11}) is a strictly monotonic increasing
function of $r$ which tends to $\alpha$ as $r\rightarrow 0$, and to
infinity as $r\rightarrow \infty$.
Consequently, no matter what choice of admissible $A$
(and thus of admissible $C$) is made, the $M_0$-spanning values of
$\bar r$ in (\ref{11}), and {\it a fortiori} in $g'_{\rm S}$, satisfy
$\bar r>\alpha$, not $\bar r>0$, and thus $g'_{\rm S}$ is not $g_{\rm DW}$\footnote
{Droste's derivation (Ref. \cite{Droste17}) differed
from that described here, since he chose to define a new radial
coordinate by setting $\overline{B}=1$, rather than by setting
$\overline{C}={\bar r}^2$. This new radial coordinate is a more complicated
function of $A$ than the right-hand side of (\ref{12}), and by
suitable choice of an integration constant its $M_0$-spanning
values can be made to satisfy $\bar r>0$. However, at the
next-to-last step of his derivation he introduced another radial
coordinate to bring the metric into the form of (\ref{6}). The
$M_0$-spanning values of this last $\bar r$ satisfy $\bar r
>\alpha$, so that his final result is not $g_{\rm DW}$, but
$g'_{\rm S}$.}. [As a matter of fact, since $g'_{\rm S}$ has the same form
as ${\bar g}_{\rm S}$, and since the $\bar r$ in (\ref{11}) has the same
$M_0$-spanning values as the $\bar r$ in (\ref{8}), $g'_{\rm S}$ is
${\bar g}_{\rm S}$.]\par
It is worth emphasizing that there is nothing wrong with making
the transformation (\ref{10}). An error arises only if it is
assumed that the resulting $\bar r$ is a ``centered'' radial
coordinate - i.e., one whose $M_0$-spanning values satisfy $\bar r
>0$ (whereas by ``noncentered'' is meant one whose $M_0$-spanning values
satisfy $\bar r>b>0$).\par
To clarify this last point, consider Minkowski's space-time
$(K,g_M)$, where in polar coordinates the components of $g_M$ are
described by (see footnote 2)
\begin{equation}\label{13}
ds^2=dt^2-dr^2-r^2d\Omega^2.
\end{equation}
If one makes the coordinate transformation
\begin{equation}\label{14}
\bar r=r+a~(a>0),
\end{equation}
then the transformed metric $(\bar g_M)$ is given by
\begin{equation}\label{15}
ds^2=dt^2-d\bar r^2-(\bar r-a)^2d\Omega^2~(\bar r>a).
\end{equation}
Again, there is nothing wrong with the use of (\ref{14}), and
$(K,\bar g_M)$ is still Minkowski's space-time, so long as one
remembers that in (\ref{15}), $\bar r=(x^2+y^2+z^2)^{1/2}+a$, so
that its $K$-spanning values satisfy $\bar r\geq a$ - that is to
say, if one remembers that values of $\bar r<a$ are meaningless,
just as are values of $r<0$ in (\ref{13}). It is only if one
decides to regard $\bar r$ in (\ref{15}) as the ordinary, centered
radial coordinate $(x^2+y^2+z^2)^{1/2}$, whose $K$-spanning values
satisfy $\bar r\geq 0$, that the interpretation of $(K,\bar g_M)$ as
Minkowski's space-time is incorrect.\par
Similarly, it is the interpretation of $\bar r$ in (\ref{6}) as
the ordinary, centered, quasipolar radial coordinate that
invalidates the derivation of $g_{\rm DW}$ from (a)-(h), and thus deprives
$g_{\rm DW}$ of the other of its two historical bases for being
interpreted as the metric of a point mass.

\section{Comparison of the Schwarzschild and Kruskal-Fronsdal
space-times}
An extension of $({\overline{M}}_0,g_{\rm DW})$ to a portion $(M_W)$ of
$R^2\times S^2$ was found by Synge \cite{Synge50},
Szekeres \cite{Szekeres60}, Kruskal \cite{Kruskal60} and
Fronsdal \cite{Fronsdal59}; the latter two also showed that
the extended space-time $(M_W,g_{\rm KF})$ contains a black hole.\par
    Because of the belief that $g_{\rm DW}$ was the metric of a point
mass, and because heretofore there has been no viable\footnote{A
Euclidean-topology but nonanalytic model was constructed by A.
Komar \cite{Komar65}; shortly thereafter
Brans (Ref. \cite{Brans65}) showed that it was afflicted with a
number of undesirable properties. A model proposed by A. Janis, E.
Newman, and J. Winicour \cite{JNW68} was
also nonanalytic, and its derivation involved a somewhat arbitrary
choice for the limit of the coefficient of $d\Omega^2$ as
$r\rightarrow {\alpha}^+$.} challenger for the role,
$(M_W,g_{\rm KF})$ has come to be accepted as {\it the} space-time of
a point mass. As seen in Secs. 3 and 4, however, neither of the
two historical bases for regarding $g_{\rm DW}$ as the metric of a
point mass is valid. Moreover, $(M_0,g_{\rm S})$ was shown in Sec. 1 to
be an eminently qualified challenger. Consequently, the point-mass
interpretation of $(M_W,g_{\rm KF})$ can no longer be based on the
fact that it is an extension of Droste and Weyl's space-time. If
such an interpretation is to be maintained, it can only be done on
the basis of its own properties - more precisely, by comparing its
properties with those of $(M_0,g_{\rm S})$, and articulating the reasons
why the former are more in accord with a point mass than the
latter. To facilitate this comparison, the following table
describes the candidates' behavior with respect to several areas
of interest:
\bigskip
\begin{table}[h]
Comparison of Schwarzschild and Kruskal-Fronsdal
Space-Times
\medskip
\begin{tabbing}
\=------------------------------------
\=------------------------\=--------------------------\kill
\>~~Property\>Schwarzschild\>Kruskal-Fronsdal\\
\\
\>1. Topology\>Euclidean\>non-Euclidean\\
\>2. Type of singularity \>point\>two-surface \cite{Ohanion76}\\
\>3. Presence of singularity\>all time\>absent for a finite time
\cite{Komar65}\\
\>4. Geometry\>globally static\>globally nonstationary \cite{Fronsdal59}\\
\>5. Curvature invariant\>finite as $r\rightarrow 0$\>infinite as
$v^2-u^2\rightarrow1^-.$
\end{tabbing}
\end{table}

Not all these properties are on the same footing as regards their
decisiveness for the choice in question. In particular, few today
would argue that the topology of a point-mass universe must be
Euclidean - about the most that can be said is that ``other things
being equal'', a simple model is preferable to a complicated one.
Likewise, there is no absolute necessity for polynomial curvature
invariants to tend to infinity as a singularity is approached
\cite{CJ74} - all that can be said is that bounded values are the
exception \cite{Clarke75}.\par
On the other hand, the behavior exhibited by Schwarzschild's
space-time with regard to 2, 3, and 4 would seem to be inherent in the
very concept of a point mass. Consequently, in order to choose
$(M_W,g_{\rm KF})$ over $(M_0,g_{\rm S})$, one should be prepared to
explain, {\it first}, how it is possible for a static phenomenon
to give rise to a time-varying geometry; {\it second}, how, in a
theory in which matter is manifested by metric singularities, it
is possible to represent a point mass for a finite time by a
universe whose metric has no singularity whatever; and {\it
third}, how it is possible for a mathematical point to be modeled
as a two-dimensional surface. Although these contradictions have
been pointed to in the past, so far as I am aware they have never
been squarely faced by the proponents of the point-mass
interpretation of $(M_W,g_{\rm KF})$. Now that the historical basis
for this interpretation has been invalidated, it appears to be
essential to do so.

\section{Conclusions}
Irrespective of which space-time is chosen, it should be clear
from the analysis of Sec. 4 that use of (\ref{11}) to represent a
spherically symmetric phenomenon generally makes $\bar r$ a
noncentered radial coordinate, which if regarded as centered will
give rise to an error similar to that which occurred in connection with
$g_{\rm DW}$. Since (\ref{11}) (and its cylindrical counterpart) has
been used to derive many of the exact solutions known today (e.g.,
the Reissner-Nordstr\"om metric\footnote{See, for example, the
derivation given by R. Tolman \cite{Tolman34}.}), the
interpretation of the associated space-times is suspect, and might
more appropriately be conferred on their ``Schwarzschild-type''
alternatives, obtained by substituting (\ref{9}) directly into the
field equations. The same suspicion also attaches to those exact
solutions which have not been derived but merely ``found'', and
which for certain values of their parameters reduce to $g_{\rm DW}$
(e.g., the Boyer-Lindquist version \cite{BL67} of Kerr's
metric).\par
If Schwarzschild's space-time should ultimately prevail as a
model for a point-mass, then two additional consequences
arise:\par
First, of course, current ideas as to the nature of physics in the
immediate neighborhood of a mass point will have to be revised,
since until now this domain has been thought to be the inside of a
black hole.\par
Second, analysis of gravitational equilibrium \cite{OV39} or collapse
\cite{OS39} involves the choice of both an ``interior'' and an
``exterior'' metric. For the spherically symmetric case the
exterior metric must be that of a point mass\footnote{For the
case of collapse to a point. In other cases, the fact that the
exterior metric need not possess property (e) permits use of a
wider class of metrics than just that of a point mass. See, e.g.
K. Schwarzschild \cite{Schw16b}.}, so that heretofore $g_{\rm DW}$ has
been used. With $(M_0,g_{\rm S})$ as the model for a point mass, the
exterior metric must be $g_{\rm S}$\footnote{For the
case of collapse to a point. In other cases, $g_{\rm S}$ should be
replaced by the generalization of the point-mass metric that is
obtained by omitting property (e). As may be seen from
Schwarzschild's paper cited in Ref. \cite{Schw16b}, this
generalization of $g_{\rm S}$ has the same form as (\ref{3}), but with
(\ref{5}) replaced by $f=(r^3+\rho^3)^{1/3}$, where $\rho$ is
independent of $\alpha$. The corresponding generalization of
(\ref{28}) is similar - (\ref{28}) is unaffected, but it is no
longer necessary that $A$ tend to zero as $r\rightarrow 0$.}, so
that all work done to date on spherically symmetric equilibrium or
collapse would be invalid. In particular, with $g_{\rm S}$ as the
exterior metric no black hole is formed no matter how far the
collapse proceeds - consequently, if uncharged, nonrotating,
spherically symmetric black holes exist, they were not created by
gravitational collapse, but are primordial.

\section*{Acknowledgments}
This paper would not have been written were it not for the
encouragement and insistence of Ray Hemann of Lockheed California
Company. Thanks are also due to F. Estabrook, R. Finkelstein, C.
Fronsdal, Ralph Smith, and H. Wahlquist for helpful discussions,
P. Bergmann, W. Bonnor, E. Newman and K. Thorne for critical
comments on an earlier version, N. Rosen for support when most
needed, and B. O'Neill and my son Howard for clarifying the
differential geometry concepts involved.

\appendix
\section*{Appendix: The most general analytic, Lorentz metric
satisfying $(\rm{a})-(\rm{h})$}
Although Schwarzschild \cite{Schw16a} did not make use of
(\ref{10}), he did impose an extraneous condition on the solution,
namely, that its determinant be $-1$ when expressed in terms of
$(t,x,y,z)$. Since this requirement has nothing to do with the
physics of a point mass, its imposition may eliminate some metrics
that are consistent with Finkelstein's (a)-(h). By dropping this
requirement and avoiding the use of (\ref{10}), the most general
solution is obtained.\par
To this end, let us suppose that $A$, $B$, $C$ denote any analytic
function of $r$ such that (\ref{9}) satisfies (a), (d), (e).
Substituting (\ref{9}) into Dingle's\footnote{See Ref.
\cite{Tolman34}, pp. 253-257.} expressions for $T^i_j$ results in
\begin{eqnarray}\label{16}
-8\pi T^1_1=-1/C+{C'}^2/4BC^2+{A'}{C'}/2ABC=0,\\\label{17}
-8\pi
T^2_2={C''}/2BC+{A''}/2AB-{C'}^2/4BC^2-{B'}{C'}/4B^2C\\\nonumber
-{A'}^2/4A^2B-{A'}{B'}/4AB^2+{A'}{C'}/4ABC=0,\\\label{18}
T^3_3=T^2_2,\\\label{19}
-8\pi T^4_4={C''}/BC-1/C-{B'}{C'}/2B^2C-{C'}^2/4BC^2=0,
\end{eqnarray}
with all other $T^i_j$ identically zero.\par
Subtracting (\ref{19}) from (\ref{16}) leads to
\begin{equation}\label{20}
2C''/C'-C'/C-B'/B-A'/A=0
\end{equation}
which integrates to
\begin{equation}\label{21}
{C'}^2=JABC,
\end{equation}
where $J$ is an arbitrary constant. Since $A$, $B$, $C$ are
positive (see Sec. 4), it follows that $J\geq 0$. But if $J$ were
zero, then ${C'}$ would be identically zero, which would make $C$
constant and thus violate property (d). Hence $J>0$, and $C'$
never vanishes.\par
Solving (\ref{21}) for $B$ and substituting therefrom for $B$ into
(\ref{16}) yields, after some algebra
\begin{equation}\label{22}
C'(C'/C+2A'/A-4C'/JAC)=0.
\end{equation}
Since $C'\neq 0$, we obtain from the other factor in (\ref{22})
\begin{equation}\label{23}
C={\alpha}^2/(4/J-A)^2,
\end{equation}
where $\alpha$ is an arbitrary nonzero constant. Substituting for
$C$ and $C'$ from (\ref{23}) into (\ref{21}) gives
\begin{equation}\label{24}
B=4{\alpha}^2{A'}^2/JA(4/J-A)^4.
\end{equation}
Substituting for $B$ and $C$ from (\ref{24}) and (\ref{23}) into
(\ref{17}) shows that the latter is satisfied identically for
arbitrary $A$. Hence, (\ref{9}) becomes
\begin{equation}\label{25}
ds^2=Adt^2-[4{\alpha}^2{A'}^2/JA(4/J-A)^4]dr^2
-[{\alpha}^2/(4/J-A)^2]d\Omega^2.
\end{equation}
Applying property (d) to the coefficient of $d\Omega^2$
shows that
\begin{equation}\label{26}
{\alpha}^2/(4/J-A)^2\sim r^2~as~r\rightarrow \infty,
\end{equation}
which reduces to
\begin{equation}\label{27}
A\sim 4/J-\alpha/r~as~r\rightarrow \infty.
\end{equation}
Applying property (d) to the coefficient of $dr^2$ adds nothing to
(\ref{27}), but applying it to the coefficient of $dt^2$ shows
that $J=4$. Hence (\ref{25}) becomes
\begin{equation}\label{28}
ds^2=Adt^2-[\alpha^2{A'}^2/A(1-A)^4]dr^2
-[\alpha^2/(1-A)^2]d\Omega^2.
\end{equation}
$A$ cannot be 1 since this would destroy the analyticity of the
coefficient of $d\Omega^2$. This, together with the positivity of
the coefficient of $dr^2$, shows that $A'$ cannot be zero.
Together with the analitycity of $A$, this means that $A'$ is
either always negative or always positive. Ruling out the former
on Newtonian grounds\footnote{$A'<0$ would require $\alpha<0$,
which in turn would give rise to negative mass when compared to
the Newtonian expression for the potential at large values of $r$
[see \cite{Levi-Civita54}]. There seems to be no way to rule this
out on the basis of (a)-(h) alone.}, it follows from (\ref{27})
that $\alpha>0$.\par
Finally, if $A\rightarrow a>0$ as $r\rightarrow 0$, then the
transformation $\bar r=A(r)$ would result in a diagonal set of
${\bar g}_{ij}$ whose elements are nonzero, finite, and have a
nonzero determinant as $r\rightarrow 0$, which would permit the
metric to be extended to $L$, contrary to property (e). Hence, $A$
must tend to zero as $r\rightarrow 0$. We conclude: The most
general analytic, Lorentz-signature metric satisfying (a)-(h) (and
having positive mass: see footnote 12) is given in quasipolar
coordinates by (\ref{28}), where $\alpha$ is an arbitrary positive
constant, and $A(r)$ is an arbitrary analytic, strictly monotonic
increasing function of $r$ which
tends to zero as $r\rightarrow 0$ and behaves like $1-\alpha/r$ as
$r\rightarrow \infty$.\par
Comparison of (\ref{28}) and (\ref{3}) shows that the $A(r)$ which
gives rise to Schwarz\-schild's solution is
\begin{equation}\label{29}
A_{\rm S}(r)=1-\alpha/f(r),
\end{equation}
which is easily seen to satisfy all of the above-mentioned
requirements on $A$. Moreover, if $A^*$ denotes any admissible
choice of $A$, the resulting metric can always be transformed into
$g_{\rm S}$ by means of the transformation
\begin{equation}\label{30}
1-\alpha/f(\bar r)=A^*(r),
\end{equation}
which converts (\ref{28}) into (\ref{3}) with $r$ replaced by
$\bar r$ - and equally important, makes $\bar r$ a strictly
monotonic increasing function of $r$, with $M_0$-spanning values
satisfying $\bar r>0$. Thus $g_{\rm S}$ can be regarded as a canonical
form for the metric of a point mass.\par
On the other hand, comparison of (\ref{28}) and (\ref{6}) shows
that the $A(r)$ which gives rise to the Droste-Weyl metric is
\begin{equation}\label{31}
A_{\rm DW}(r)=1-\alpha/r,
\end{equation}
which violates the requirement that $A\rightarrow 0$ as
$r\rightarrow 0$.\par
Finally, the simplest metric obtainable is that corresponding to
$A(r)=r/(r+\alpha)$. This gives
\begin{equation}\label{32}
ds^2=[r/(r+\alpha)]dt^2-[(r+\alpha)/r]dr^2-(r+\alpha)^2d\Omega^2.
\end{equation}

\vbox to 2.0cm{}

\section*{Erratum \cite{Abrams80a}}
1st and 3rd lines under Eq. (\ref{6}): Replace $M_0$ by
$\overline{M}_0$.\par
9th line under Eq. (\ref{8}): Insert ``, when regarded as a
mapping from $M$ to $M$,'' between ``transformation'' and
``be''.\par
2nd line under Eq. (\ref{11}): Replace $A$ by $\overline{A}$,
and $B$ by $\overline{B}$.\par
Also, the proof of the requirement $A(0^+)=0$ is incorrect;
however, the requirement is still valid \cite{Abrams80b}.

\newpage
\bibliographystyle{amsplain}

\end{document}